# Unraveling the degradation mechanism in FIrpic based Blue OLEDs: II. Trap and detect molecules at the interfaces

Marta Penconi[a], Marco Cazzaniga[a]†, Walter Panzeri[b], Andrea Mele[b,c], Fausto Cargnoni[a*], Davide Ceresoli[a*], Alberto Bossi[a*]

[a] Istituto di Scienze e Tecnologie Molecolari - CNR, PST di via Fantoli 16/15, 20138 Milan and via C. Golgi 19, 20133 Milano, Italy; SmartMatLab Center, Via C. Golgi 19, 20133 Milano, Italy, [b] Istituto di Chimica del Riconoscimento Molecolare - CNR, "U.O.S. Milano Politecnico" Via Mancinelli 7, 20131 Milano Italy, [c] Dipartimento di Chimica, Materiali e Ingegneria Chimica "Giulio Natta", Politecnico di Milano, Via Mancinelli, 7, 20131 Milano, Italy

**ABSTRACT:** The impact of organic light emitting diodes (OLEDs) in modern life is witnessed by their wide employment in full-color, energy-saving, flat panel displays and smart-screens; a bright future is likewise expected in the field of solid state lighting. Cyclometalated iridium complexes are the most used phosphorescent emitters in OLEDs due to their widely tunable photophysical properties and their versatile synthesis. Blue-emitting OLEDs, suffer from intrinsic instability issues hampering their long term stability. Backed by computational studies, in this work we studied the *sky-blue* emitter **FIrpic** in both *ex-situ* and *in-situ* degradation experiments combining complementary, mutually independent, experiments including chemical metathesis reactions, in liquid phase and solid state, thermal and spectroscopic studies and LC-MS investigations. We developed a straightforward protocol to evaluate the degradation pathways in iridium complexes, finding that FIrpic degrades through the loss of the picolinate ancillary ligand. The resulting iridium fragment was than efficiently trapped "*in-situ*" as BPhen derivative **1**. This process is found to be well mirrored when a suitably engineered, FIrpic-based, OLED is operated and aged. In this paper we (i) describe how it is possible to effectively study OLED materials with a small set of readily accessible experiments and (ii) evidence the central role of host matrix in trapping experiments.

INTRODUCTION

Organic electronic technologies, which include light emitting diodes, field effect transistors and organic photovoltaics, are vibrant research sectors from both the academic and industrial standpoint.[1] In particular organic light emitting diodes (OLEDs) stand out like the reference technology employed in energy saving flat panel displays and looked as promising alternative to develop solid state lighting.[2] These devices consist of organic and organometallic materials arranged in a thin-film multilayer architecture, either solution-processed or vacuum-deposited, sandwiched between two electrodes. In OLEDs, light is generated and emitted as the result of the electroluminescence (EL) process which involves i) injection of electrons and holes under applied voltage from the cathode and anode, respectively, ii) migration of the charges through the device layers and iii) charge trapping and recombination, with formation of excitons, at the emitting material. Excitons will ultimately radiatively decay to the ground state. Phosphorescent transition metal complexes (mostly based on Ir(III) and Pt(II) metal ions), doped in a host material, have been shown to yield 100% internal quantum efficiency by harvesting both singlet and triplet electrogenerated excitons.[3] High external quantum efficiencies have been achieved combining light extraction techniques with device engineering. Improved exciton recombination in the emissive layer has been obtained using blocking layers and carefully selecting the layered materials to improve the charge injection and charge transport balance. Cyclometalated iridium complexes are the most used phosphorescent emitters because of their fine tunable photophysical properties and the versatile synthesis.[3,4] Iridium complexes have been developed to span emissions from the deep UV down to the near IR spectral region as a result of the great demand for commercial applications. Yet, two big issues are evident: the lack of efficient emitters for the NIR region[3] and the severe instability issues of blue emitters.

Especially in the latter contest, the study of the degradation processes, ultimately leading to OLED failure, is essential to meet the device lifetime and the color stability needed for commercialization. Degradations in OLEDs can occur through extrinsic mechanisms as well as through a long-term intrinsic degradation.[5,6] Extrinsic mechanisms, such as the formation of non-emissive dark spots and electrical shorts, are tightly linked to the fabrication conditions. A careful control and optimization of the processing methods can ultimately limit and avoid their occurrence; for instance, high-vacuum thermal evaporation methods offer the best conditions to meet the standards of efficiency and lifetime of commercial devices.[7-9] On the contrary, the intrinsic mechanisms, still holding many open issues, mainly consist in the chemical alteration of the organic or metal-organic materials affecting the device emission efficiency (*i.e.* brightness decrease and operating voltage rise),[10] and they are particularly relevant in the emitting layer. While a recent work highlights the role of the exciton-induced generation of radical ion pairs between the host and the dopant emitter,[11] we are mainly interested in the fate of the exciton once it is trapped on the dopant emitter and it can initiate intramolecular degradations from the molecular excited state (therefore forming nonradiative recombination centers and luminescence quenchers). The understanding of chemical degradations occurring in phosphorescent blue-emitters is far more important compared to degradation of the green or red ones. In fact, the higher energy of the emitted blue photons,[4] *i.e.* the higher energy of the emitter excited state, combined to the long (~0.5-5 µs) triplet lifetime, can lead to the thermal population of non-radiative excited states (involving antibonding orbitals). As a result, blue OLEDs display shorter lifetime.

The study of emitter degradation is challenging and requires the *a priori* identification of the decomposition products to frame a mechanistic understanding. The difficulties, from the experimental standpoint, lie in the fact that emitters are doped in low concentration in the EML and only a small fraction of them degrade; on one hand, even tiny amount of degraded compounds produce efficient traps, on the other hand, very small quantity of material is available for analysis (considering the average of 20-40 nm thickness of the emissive layer), limiting the number of suitable analytical techniques. The high-performance liquid chromatography coupled with UV-Vis[12] and mass spectrometer (HPLC-MS)[13] detectors or laser desorption/ionization time-of-flight mass spectrometry (LDI-TOF-MS)[14,15] also combined with X-ray photoelectron spectroscopy (XPS)[16] offer the highest sensitivity for such studies and have been employed in early investigations. Several studies reported ligand dissociation reaction under device operation and demonstrated that it can be an irreversible process if the resulting coordinatively unsaturated fragment undergo further reactions with surrounding molecules from the adjacent transport layers, for example 4,7-diphenyl-1,10-phenanthroline (BPhen). This has been investigated by Leo *et al.* for a variety of organometallic iridium complexes such as $Ir(MDQ)_2(acac)$,[17] $Ir(ppy)_3$[18,19] and FIrpic[20,21] employing LDI-TOF-MS technique, although a possible damage effect of the incident laser during the analysis cannot be completely ruled out as source of fragmentation.[5,16] Moreover, evidences of red shift in the electroluminescence spectra under device operation have been reported, whether ascribed to ligand substitution[22,23] or related to emitter aggregation (FIrpic).[9,24] Thompson and co-worker reported on the device aging and photochemical stability of the well-known $Ir(ppz)_3$.[23] This is also the case of FIrpic (bis[2-(4,6-difluorophenyl)pyridyl-C2,N](picolinato) iridium(III)) (Scheme 1), the most representative blue phosphorescent emitter. It provides highly efficient blue OLED as dopant in host matrix but it is highly instable leading to fast failure in operating device.[25] Several studies reported that ligand dissociation in FIrpic can be irreversible since it can be followed by $CO_2$ loss.[20] Moreover, the cleavage of fluorine atoms has been observed after thermal evaporation of the material and device aging.[26]

Cargnoni and Ceresoli thoroughly detailed the degradation mechanism of FIrpic in both ground and excited states and by exciton formation.[27] Among their results, it stands out the weakness of the picolinate ligand (*pic*) towards excited state dissociation. In summary, FIrpic sequentially undergoes a bond breaking at the Ir-N(*pic*) site followed by loss of the picolinate ligand.[27] In this hypothesis, the coordinative open, left-over, Ir fragment could be trapped with proper molecules. On this ground, our experiments are designed to track the degradation follow-up. In this study we speculated, thought later proved, that once picolinate falls apart (Scheme 1), the degraded molecule could be trapped by a strongly coordinating bidentate phenantroline ligand, BPhen, thus obtaining the charged heteroleptic $[Ir(F_2ppy)_2(BPhen)]^+$, complex **1**. The simplified scheme of the overall degradation/trapping process is depicted in Scheme 1. The use of BPhen provides complex **1** with properties significantly different compared to FIrpic allowing for multi-technique detection/analysis; in particular, complex **1** is a charged compound, its molecular weight is different from that of the FIrpic and in addition it is reported to have orange luminescence.[28]

This work aims at presenting a critical experimental investigation of the FIrpic degradation process through studies encompassing the intrinsic chemical lability of the FIrpic ligands, the optical properties of FIrpic and derivative **1**, and finally the degradation environment inside an OLED under electrical stress (*i.e.* promoting degradation under electrically pumped exciton formation). Therefore, the next sections are organized as follow: 1) synthesis of the complex **1**, its optical properties and those of FIrpic; 2) FIrpic reactivity towards thermal ligand metathesis reaction with BPhen, both in solution and in solid phase, and development of suitable analytical and optical frameworks; 3) design of engineered OLED with a bulk interface between FIrpic and BPhen (the latter used also as electron transport layer - ETL), aging and real time monitoring of device degradations under electrical stress; 4) morphological, spectroscopic and chemical analysis on OLED devices. For its conception, the experimental set up is worth for a broader exploitation in OLED material analysis and would provide guidelines in the molecular engineering of stable phosphorescent emitter.

RESULT AND DISCUSSION

*1. Synthesis of the complex **1**, optical properties of complex **1** and FIrpic*

According to the theoretical framework, that suggests the detachment of the picolinate ligand from FIrpic,[27] the fate of the emitter can be followed by inducing the coordination of the BPhen molecule to the unsaturated residue [Ir(F$_2$ppy)$_2$]$^+$ giving the cationic complex **1**, as depicted in Scheme 1. For comparative purpose, complex **1** has been synthesized according to literature procedure (see supporting information for details).[28] In Figure 1a, the absorption and emission spectra of the two iridium complexes, FIrpic and **1**, are reported and their photophysical properties are summarized in Table S1. In deaerated CH$_2$Cl$_2$ solution at room temperature, FIrpic displays a structured emission with maximum at 468 nm, arising from a mixed LC-MLCT character of the transition.[24] It has phosphorescence quantum yield of 97% and a lifetime of 1.7 μs.

In comparison, the emission spectrum of complex **1** is significantly red shifted with peak maxima at 514 nm. The phosphorescence quantum yield and lifetime are 51% and 5.4 μs, respectively. In this case, a dominant MLCT character of the emission is suggested by the slightly broader and featureless phosphorescence. When deposited as neat thin film by spin coating, the emission of both complexes are red-shifted, with maxima at 481 nm and 548 nm for FIrpic and complex **1**, respectively (Figure 1b); whereas FIrpic still shows a partial structured emission, complex **1** has a broad gaussian profile.

**Scheme 1. Ligand dissociation process of FIrpic and coordination to BPhen molecule to form complex 1.**

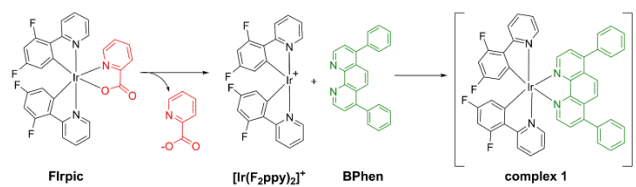

The quantum yields and lifetimes in neat thin film are respectively 15% and 0.16 μs for FIrpic and 37% and 2.8 μs for complex **1**.

*2. Thermal ligand metathesis reactions*

In order to prove the dominant mechanism of picolinate loss in FIrpic, its reactivity towards thermal ligand metathesis with BPhen was explored in solution and also in the solid state; the latter mirroring the conditions of the stacks in a OLED device. An equimolar solution of FIrpic and BPhen in argon degassed 2-ethoxyethanol was heated at 100 °C for few hours. The reaction, monitored via thin layer chromatography, shows the slow disappearance of FIrpic and the formation of a new brightly luminescent compound with lower Rf, signature of the more polar/ionic nature of the newly formed compound (as expected by the replacement of a monoanionic ligand – *i.e.* F$_2$ppy or *pic* - with the neutral BPhen). Electrospray mass spectroscopy (ESI-MS) analysis were carried out to monitor the crude reaction mixture after 3, 6 and 24 hours. Figure 2 displays the ESI-MS spectra after 1 day; the base peak at *m/z* = 905 and the isotopic pattern are consistent with the ion of elemental composition [C$_{46}$H$_{28}$N$_4$F$_4$Ir]$^+$, assignable to complex **1**. Noticeable, no co-products deriving from metathesis of BPhen with ligand other than the *pic* were found (*i.e.* excluding the abstraction of F$_2$ppy), as well as no significant amount of residual FIrpic. A freshly prepared FIrpic-BPhen mixture was injected as a control to rule out any artefact deriving from gas phase reaction inside the ESI source; the ESI analysis (Figure S4) shows the protonated or sodium cationized species of the BPhen and FIrpic as well as a possible clustered cationic system of [FIrpic-BPhen-Na]$^+$ at m/z 1050, while no evidence of the signal at m/z 905 has been observed.

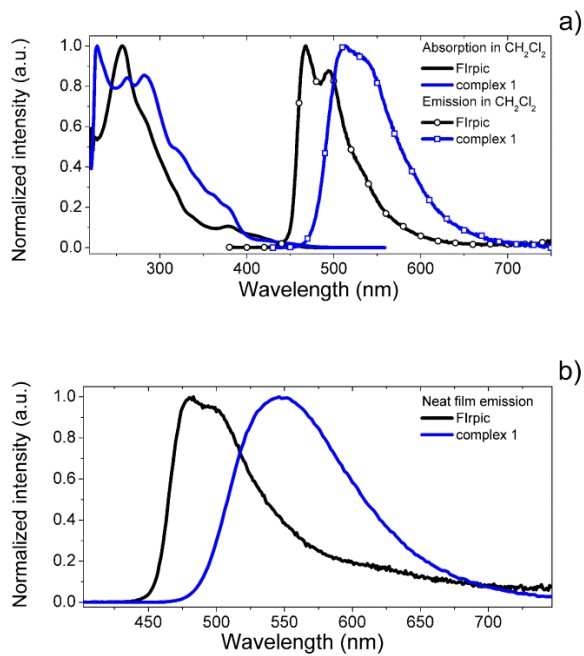

Figure 1. a) UV/Vis absorption (straight line) and emission (line with drawings) normalized spectra in deaerated $CH_2Cl_2$ solution and b) normalized emission spectra of neat thin films of FIrpic (black) and complex **1** (blue).

Hence, the reactivity of an equimolar blend of FIrpic and BPhen was investigated in solid state. Thermal analysis was followed via differential scanning calorimetry (DSC) in the temperature range up to 300 °C ($N_2$ flushed, open vessel). Upper limit temperature was chosen according to the melting point of BPhen (218-220°C) and that of FIrpic (330-335°C). DSC was run either placing the finely grinded powder mixture of the two compounds in the DSC crucible (experiment DSC-A) or dissolving the mixture in dichloromethane and let the solution evaporate in the crucible (experiment DSC-B). The DSC curves are reported in Figure S5. In both cases, a broad exothermic peak appears at 250°C with onset around 180°C. DSC-A showed the endothermic melt of BPhen at ca. 220°C immediately prior the exothermic reaction. In the second case (DSC-B), a slight melting is only perceivable around 180°C followed by exothermic degradation. The residue in the crucibles were than subjected to HPLC-MS analysis and compared to the as-prepared FIrpic-BPhen mixture. After both DSC-A and DSC-B experiments, the residue analysis pointed out the disappearance of FIrpic signal and the presence of **1** (see Figures S6, S7 and S8).

The solid state reactivity was also followed by photoluminescence (PL) experiments. Thin films with similar architecture as later used in the OLEDs were prepared by spin coating. A FIrpic-BPhen neat film (1 to 1 ratio) was spin-coated onto a quartz slide and then thermally annealed for 30 min at 100°C.

The photoluminescence spectrum of the pristine film, peaked at 470 nm, evolves upon annealing to a broad peak with maximum at 560 nm grown as a shoulder to the FIrpic emission (Figure 2); this band matches the solid state luminescence of **1**. Proving that the spectral change is the consequence of the FIrpic/BPhen metathesis reaction, PL upon thermal annealing was investigated both in the neat film of FIrpic as well as for a FIrpic-BPhen pair dispersed into a PMMA matrix (4%w/w), showing, in both cases, no significant spectral alteration (Figures 2 and S9).

Altogether, formation of complex **1** is confirmed both by ESI-MS analysis and PL experiments; the two techniques provide accessible and independent methods to follow the metathesis process. Mass spectrometry can discriminate FIrpic from **1** on the basis of an over 200 Da difference in molecular weight, likewise their emission wavelengths differ of 50 - 70 nm (respectively for the solution and the neat thin film). The trapping reaction is efficient both in solution and, worth standing, in solid state when the two compounds are intimately mixed; DSC and PL prove (i) the formation of **1** as the single degradation/trapped product, (ii) the possibility to follow the process by mass-spectrometry and photoluminescence, and (iii) the occurrence of an intermolecular reaction to form **1** when the two molecule are in close proximity.

3. *Engineered OLED with an interface between FIrpic and the chemical trap BPhen*

Under operation, OLEDs would provide suitable conditions for the emitter degradation through the electrical generation of excitons (rather than by thermal mechanism). In particular, according to the theoretical framework, the emitter is even more prone towards ligand dissociation when triplet exciton are formed, given the very low activation barrier for Ir-*pic* detachment.[27] Four architectures of FIrpic based devices (Figures 3 and S10) were prepared by physical vapor deposition (PVD). With the aim to promote efficient trapping of the FIrpic degra dation product by coordination to BPhen, the emitter is employed as neat layer sandwiched between the common hole transport layer (HTL) N,N'-di[(1-naphthyl)-N,N'-diphenyl]-1,1'-biphenyl)-4,4'-diamine (NPD) and BPhen (used as ETL). This type of device, with the simplest architecture ITO/NPD/FIrpic/BPhen/LiF/Al, was of limited utility in this study due to significant voltage dependent electroluminescence (Figure S10), caused by electrons leaking into the NPD layer (as evidenced by the growth of the 450 nm band at high driving potentials). In order to confine electrons into the emitting/reacting FIrpic/BPhen interphase and generate excitons on the FIrpic molecules, $Ir(ppz)_3$ (*fac*-tris-iridium(III)-(1-phenylpyrazole)) was introduced as electron blocking layer (EBL) between the NPD and FIrpic ones.[29]

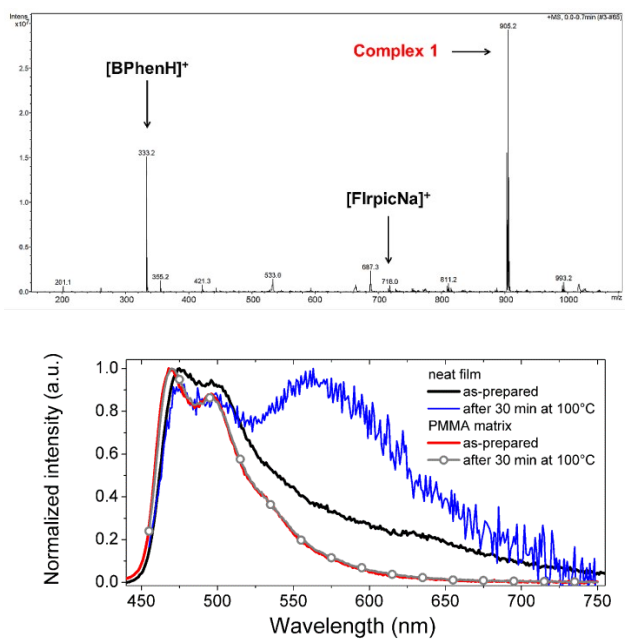

Figure 2. (top) Positive polarity ESI-MS spectrum of the crude thermal metathesis reaction mixture FIrpic-BPhen after 1 day. $MW_{FIrpic}$=695.1 g/mol, $MW_{BPhen}$=332.1 g/mol, $MW_1$=905.2 g/mol. (bottom) Normalized emission spectra of FIrpic-BPhen mixture in neat film before (black line) and after (blue line) thermal treatment at 100°C for 30 min and in PMMA matrix before (red line) and after (grey line with drawings) thermal treatment, $\lambda_{ex}$=430 nm.

Two device architectures were prepared, the first one having a planar junction configuration with the FIrpic layer adjacent to the BPhen one (device A in Figure 3); the second device having a bulk heterojunction layer of the two materials co-evaporated in 1 to 1 ratio and sandwiched between the neat layers of the two compounds (configuration B in Figure 3, namely devices $B_{20}$ and $B_{80}$, the two differ in the thickness of the heterojunction layer, been respectively 20 and 80 nm).

A third architecture was built as a reference, where FIrpic was doped into N,N'-dicarbazolyl-3,5-benzene (mCP) matrix (device C in Figure 3).[30] Figures 4, 5 and 6 report the energy levels and layer thicknesses of the materials in the respective stacks, the electroluminescent spectrum before and after aging, and the I/V/EL$_{intensity}$ profiles. Voltage independent EL is obtained, indicating the correct charge trapping and exciton confinement in the EML.

Device A switches on at a $V_{on}$ = 4 - 5 V. The initial EL spectrum perfectly matches the PL one of FIrpic neat film (Figure 4b). The device was then aged at constant current density (160 mA/cm$^2$) for 50 min. The EL intensity drops to 17% of the initial value after only 10 min of electrical aging and continues to decrease as time elapses. The EL spectrum after aging shows a remarkable change: the FIrpic emission decreases, becoming almost negligible, while a new band at 580 nm appears. Visually, the color of the emitted light changes from blue to green-yellow and the brightness noticeably drops. The current density behavior versus the applied potential decreases by four times at the end of the electrical stress (Figure 4c). Analogously, the aging at 160 mA/cm$^2$ for 50 min applied to the device $B_{20}$ (Figure 5) results in a drastic decrease of the EL intensity (being 20% of the initial value after 10 min); the FIrpic emission is quickly replaced by a broad band centered 550 nm, again resembling the luminescence of the complex **1**. The I/V curve of the device before and after aging shows the same behavior observed for the device A. In both devices, the prominent EL redshift reproduces very nicely the spectral shift observed in the thermal experiments described above (section 2).

As a control, device C was made in a typical doped architecture with FIrpic dispersed in 6% weight into the mCP host. Figure 6 reports the energetics, thickness and properties of the device. Prior aging the EL spectrum matches the PL emission of FIrpic in solution (Figure 6b) while driving the device at 160 mA/cm$^2$ or even at 16 mA/cm$^2$, results in a remarkable decrease of both EL intensity (over 90% of intensity is lost after 10 min at 160mA/cm$^2$) and current density in similar manner as in devices A and B. On the other hand, its EL spectrum shows no significant modification given the lack of any macroscopic contact between the FIrpic and the BPhen ETL; the loss of efficiency would arise from exciton induced formation of dark traps originated by FIrpic degradation. This observation fits with the experiment carried out dispersing FIrpic and BPhen in PMMA (section 2).

Taken together these results confirm that only in the presence of a macroscopic interface between FIrpic and BPhen (the bilayer heterojunction or the bulk heterojunction), the exciton induced degradation of FIrpic is quickly sidestepped by the efficient entrapment of coordinatively unsaturated Ir system by the BPhen molecule in close contact. Results in line with the solid state behavior (section 2) upon thermal annealing.

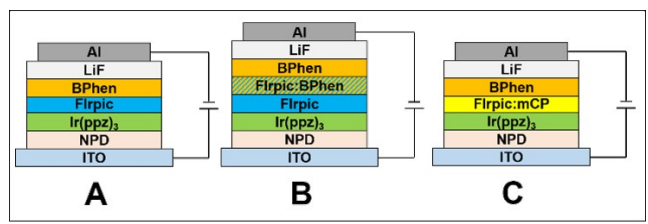

Figure 3. Configuration of the devices with planar FIrpic-BPhen interface (device A), bulk hererojunction (device B) and FIrpic doped into mCP matrix (device C).

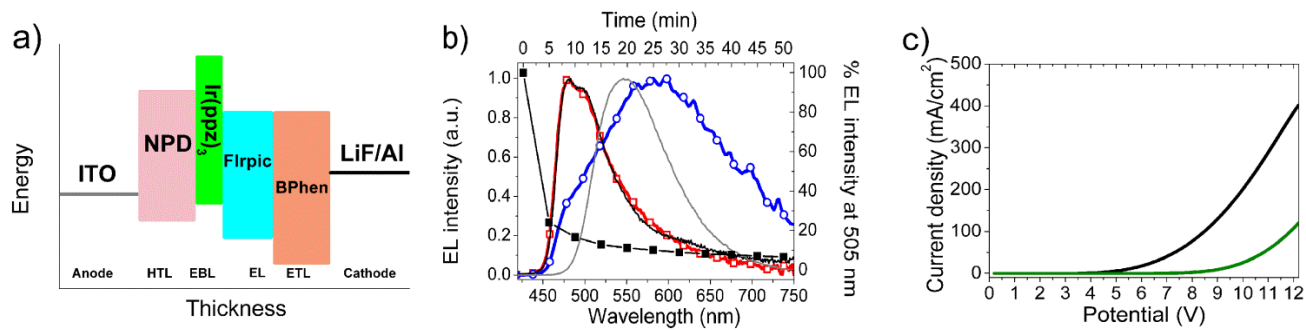

Figure 4. a) Energy diagram for the device A: ITO/NPD(50nm)/Ir(ppz)$_3$(10nm)/FIrpic(20nm)/BPhen(50nm)/LiF(1nm) /Al(80nm). b) Normalized electroluminescence of the unaged device (red line with squares) and of the device driven at 160 mA/cm$^2$ for 50 minutes (blue line with circles) compared with the photoluminescence of FIrpic (black) and complex **1** (grey) in neat film; reduced intensity of electroluminescence versus time during aging. c) Current density-voltage curve of the unaged (black) and aged (green) device.

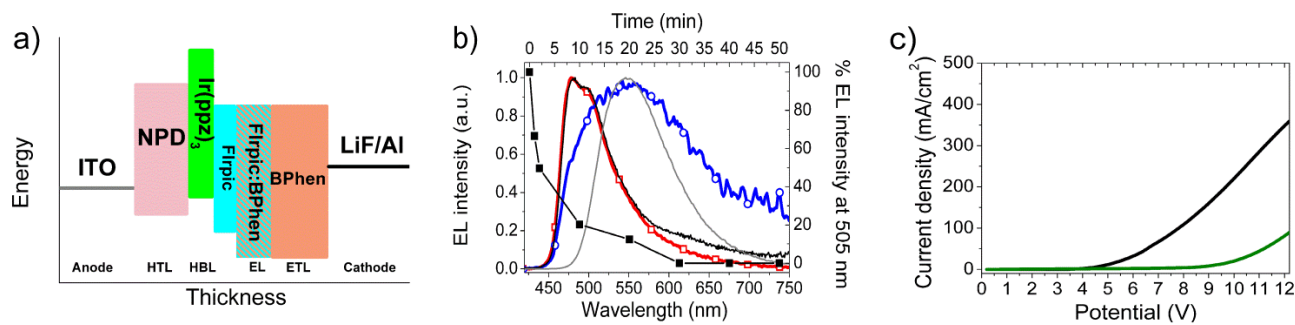

Figure 5. a) Energy diagram for the device B$_{20}$: ITO/NPD(50nm)/Ir(ppz)$_3$(10nm)/FIrpic(10nm)/ FIrpic:BPhen 50% mol (20nm)/BPhen(40nm)/LiF(1nm)/Al(70nm). b) Normalized electroluminescence of the unaged device (red line with squares) and of the device driven at 160 mA/cm$^2$ for 50 minutes (blue line with circles) compared with the photoluminescence of FIrpic (black) and complex **1** (grey) in neat film, reduced intensity of electroluminescence versus time during aging. c) Current density-voltage curve of the unaged (black) and aged (green) device.

### 4. *Morphological, spectroscopic and chemical analysis on OLED devices*

Although the EL signature in the aged devices A and B are in optimal agreement with the degradation-entrapment mechanism described, a direct evidence of the formation of **1** in aged OLEDs would be more assertive and conclusive. We therefore employed PL and ESI-MS analysis to study the aging processes. We fabricate at least two identical devices (for each configuration, A and B) in a single chamber load; one was kept as benchmark without being electrically aged while the second was aged according the described procedure.

To mimic the FIrpic/BPhen interface an incomplete device A was built up to the FIrpic deposition, the morphology and topography of outermost layer (Figure S12) was analyzed by non-contact Atomic Force Microscopy (AFM). The surface appears, overall, quite flat with an average hill-to-valley roughness of 4 nm. Nonetheless the FIrpic layer is characterized by depressions where more than 60% of the mapped-area have depth greater that the average roughness. An hill-to-valley distances even exceeding 10 nm are found on the surface implying that the real FIrpic/BPhen interfacial contact area is much larger than the nominal pixel area.

Full device A was thus studied by PL experiments applying on the device a positive mask with a 6.25mm$^2$ aperture on the probed pixel. Without any change in device position and instrumental optical setup, we recorded the unaged pixel emission (excitation at 425 nm). The pixel was than aged according to the described protocol and PL spectrum was measured again. Although the EL changes as in Figure 4b, the PL spectrum is only slightly changed (Figure S13); indeed when we subtracted the normalized unaged pixel PL spectrum from that of its aged one, surprisingly we observed a bleach of the FIrpic emission at 460 nm and the raise of the emission of **1** at 550 nm. PL emission was also measured by applying to the aged device a bias voltage at which no EL can be detected, namely 5 V forward bias and 5 V reverse bias. The corresponding PL spectra are identical to the one in absence of bias, indicating that **1** is formed irreversibly at the FIrpic/BPhen interface.

For the ESI-MS chemical analysis, pixels of device B$_{20}$ were aged as described in section 3. The cathode was tape-peeled[31] and the underneath organics of the pixels were dissolved in 1 drop of acetonitrile then diluted into 100 μL of the same solvent. At first, the samples were subject to a direct-infusion ESI-MS scan. The mass spectra of the unaged and aged devices (Figure S14) show overall very similar pattern in which all the chemicals used in the fabrication can be readily identified; the sample obtained from the aged device displays also few other signals, possibly due to the contaminations from the tape-pealing process instead of being correlated to the possible matrix alteration described elsewhere.[32] Yet this analysis does not provide the striking evidence for the presence of complex **1** as the intensity of $m/z = 905$ is at the noise level).

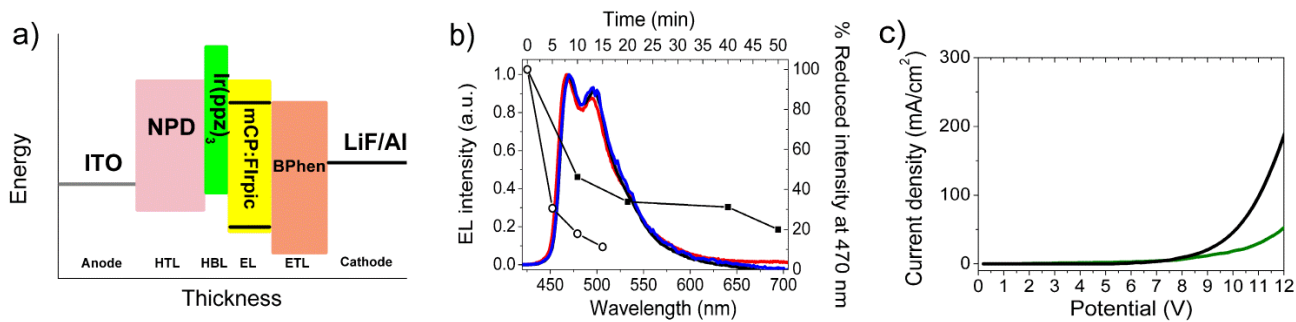

Figure 6. a) Energy diagram for the device C: ITO/NPD(50nm)/Ir(ppz)$_3$(10nm)/FIrpic:mCP(6%wt, 30nm)/BPhen(40nm) /LiF(1nm)/ Al(75nm). b) Normalized electroluminescence of the unaged device (black) and of the device driven at 16 mA/cm$^2$ for 50 minutes (blue) compared with the photoluminescence of FIrpic in deaerated dichloromethane solution (red); reduced intensity of electroluminescence versus time during aging at 16 mA/cm$^2$ (filled squares) and 160 mA/cm$^2$ (open circles). c) Current density-voltage curve of the unaged (black) and aged (green) device.

In order to avoid possible artefacts due to the direct-infusion of a "multi-component" mixture and reduce the total ionic load of the MS-detector with compounds at concentration order of magnitudes higher than the one expected for **1** (see below),[33] we developed a 20 minutes reverse-phase gradient HPLC method. This method allows to perfectly separate all the organic (BPhen and NPD) and organometallic (FIrpic, Ir(ppz)$_3$) components in the OLED stacks as well as to elute complex **1** in a *clean* region of the chromatogram (at ca 9-10 min). In Figure S15 are reported the HPLC chromatograms of the compounds mixture as well as the ones of all the chemicals separately. We used the retention times to locate complex **1** in the HPLC profile and confirm its presence / absence in aged and unaged samples via MS analysis. The first HPLC-MS study performed on the device B$_{20}$ was, in this case, of limited utility given the low S/N ratio and the presence of background $m/z$ 905 signal; the whole (ESI)MS spectra extracted from the HPLC run at the retention time of complex **1** reveal in both the unaged and aged pixels (Figure S16) presence of variable amount of **1**. A new attempt to prove the *in-situ* formation of **1** upon aging was done replicating the previous analysis on the prototype device $B_{80}$. Given the four times thicker bulk heterojunction emissive layer, the formation of a significantly larger amount of **1** was expected. In addition the MS spectrometer was set mass isolation mode on *m/z* 905 in order to achieve the highest instrumental sensitivity. The mass chromatogram of the unaged pixels of the isolated *m/z* 905 signal gave an experimental intensity larger than expected for a "fresh" system. The integrated area at the retention time of ca 10 min, was $8.9 \cdot 10^6$ units; the corresponding mass peak showed the expected isotopic pattern for **1**. Although surprising this finding is not unexpected, given the recent observation of Forrest *et al.*.[16] In the case of the aged $B_{80}$ pixels, the integrated intensity of the same peak was as large as $15.1 \cdot 10^6$ units; in addition a new chromatographic peak at 10.8 min (integration $1.82 \cdot 10^6$ units) appeared with the same isotopic pattern of the peak at 10 min. It is worth pointing out that such signal, <u>appearing</u> only in the aged device and at slightly <u>longer retention time,</u> can be assigned to a "facial-like" isomer of **1,** resulting from the loss of the *trans*-arrangements of the F2ppy ligands in the Ir fragment before the trapping event (figure S17 and 18).[34]

In order to finally establish whether the area of the mass chromatogram of **1** in the aged pixels is significantly larger than that of the same signal in the unaged device, we compared the NPD signal intensity on the integrated UV chromatograms for the two experiments. NPD is selected as internal standard because is the bottom layer in the device. In the sample preparation, it is dissolved after all the other over-layers are necessarily dissolved. The NPD area in unaged $B_{80}$ is 238 arbitrary units while in the aged one is 290. In the assumption that the aging process is not affecting the elute composition, and set to 1 the intensity of the chromatographic peak of isolated 905 in unaged device, the *m/z* 905 peak in the aged one is 1.6 (table S2), that is considered a significantly higher value suitable to demonstrate the degradation–trapping process proposed in this work.

## CONCLUSION

In this study we have set up simple and selective trapping experiments proved to be effective in simulating the *ex-situ* and *in-situ* degradation of FIrpic. We developed a combination of complementary, mutually independent analytical techniques, including DSC, HPLC/MS and PL experiments easily accessible to any chemistry lab, to follow this process. We safely demonstrated that the results of these "chemical studies" are indeed well mirrored when suitable OLED prototypes are fabricated and aged.

We investigated the reactivity of FIrpic in the ground state towards ligand substitution with BPhen by thermal reaction, as alternative to photo-induced ligand metathesis reaction. The formation of complex **1** was ascertained in solution and, most importantly, in solid state. This allowed to elucidate in a simple and easily reproducible way the occurrence of degradation-trapping in working OLED devices.

Two engineered OLED architectures, characterized by the presence of a planar or bulk interface between FIrpic and BPhen, were designed. Upon electrical aging, the electroluminescence spectra evolve from the typical bluish emission of FIrpic to a broad orange color that fits with emission of complex **1**. This color change perfectly matches the PL color evolution of a thin film FIrpic-BPhen blend upon thermal treatment. In these experiments, the raise of the orange band is indisputably associated with formation of complex **1** as proved by ESI-MS analysis. By means of PL difference spectra we were able to probe that on the aged pixel, it is possible to observe the raise of the emission of **1** in lieu of the bleach of the FIrpic one. Most importantly, applying HPLC-MS techniques with the isolation of the target molecular ion (*m/z* 905) it has been undisputable proved the formation of the complex **1** as the result of a trapping event between the degraded FIrpic and the BPhen trap. Interestingly in this case it was possible to discriminate in the MS chromatogram two peaks with 905 *m/z* belonging to two geometrical isomers

We believe that these experiments have provided an easy reproducible framework to circumvent the problems linked to trace analysis of the formation of degradation product in OLEDs and can be applied even in a feedback scheme for a simple evaluation of the stability of new emissive dopants.

It is worth mentioning that these results also point to the fact that if a host matrix is designed to efficiently trap the degradation intermediates into high energy derivative which will not funnel the excitonic energy of FIrpic, this would lead to a new paradigm to extend the device lifetime.

## EXPERIMENTAL SECTION

Sublimated grade 4,7-diphenyl-1,10-phenanthroline (BPhen), N,N′-di[(1-naphthyl)-N,N

'-diphenyl]-1,1'-biphenyl)-4,4'-diamine (NPD) and N,N'-dicarbazolyl-3,5-benzene (mCP) were purchased from TCI, bis[2-(4,6-difluorophenyl)pyridyl-C2,N](picolinato)iridium(III) (FIrpic) and LiF were purchased from Aldrich, Al pellets was purchased from Kurt J. Lesker Company. *Fac*-Ir(ppz)$_3$ and complex **1**, [Ir(F$_2$ppy)$_2$(BPhen)]PF$_6$, were synthesized according to literature procedures.[28] from commercially available 2-(2',4'-difluorophenylpyridine) and 1-phenylpyrazole purchased from TCI *fac*-Ir(ppz)$_3$ was sublimated before use. All solvents are degassed under argon prior the use.

NMR spectra were recorded at 300 K on Bruker AV400 and Bruker AC300 spectrometers. Chemical shifts are expressed in parts per million (ppm) and referred to the residual signal of the solvent (2.50 ppm in DMSO-d6).); coupling constants are given in Hz.

Reverse phase gradient HPLC analysis were performed on Agilent 1100 equipped with VWD UV-Vis detector (set 300 nm), a Waters Atlantis column (4.6 mm x 10.0 cm, 3 μm particle size) and a mass spectrometer BRUKER Esquire 3000 PLUS with quadrupole ion-trap detector. HPLC analysis ran in a 20 min gradient elution, flow 0.5 ml/min starting from 70-30 acetonitrile – water mixture up to 100% acetonitrile in the presence of 0.05% trifluoroacetic acid (TFA). Direct infusion electrospray ionization mass spectrometry (ESI-MS) experiments were carried out on the same mass spectrometer. The samples were prepared by diluting suitable stock solutions of the substrates in CH$_3$CN up to an overall concentration of $10^{-6}$ M. The solutions were infused directly in the ESI source at 4 μL min$^{-1}$.

UV/Vis absorption spectra were obtained on Shimadzu UV-Vis-NIR Spectrophotometer in 1 cm path length quartz cell with dichloromethane. Photoluminescence quantum yields were measured with a C11347 Quantaurus - QY Absolute Photoluminescence Quantum Yield Spectrometer (Hamamatsu Photonics U.K), equipped with a 150 W Xenon lamp, an integrating sphere and a multi-channel detector. Photoluminescence emission spectra, lifetimes and electroluminescence spectra were obtained with a FLS 980 spectrofluorimeter (Edinburg Instrument Ltd.). Continuous excitation for the steady state measurements was provided by a 450 W Xenon arc lamp. Emission spectra were corrected for the detector sensitivity. Photoluminescence lifetime measurements, determined by TCSPC (time-correlated single-photon counting) method, were performed using an Edinburgh Picosecond Pulsed Diode Laser EPL-375 (Edinburg Instrument Ltd.), with central wavelength 374.0 nm, pulse width 66.2 ps and repetition rates 50 μs. Photoluminescence experiments at room temperature were carried out in nitrogen degassed dichloromethane solution $2 \times 10^{-5}$ mol L$^{-1}$. Current-voltage measurements were obtained from AUTOLAB PGSTAT302N equipped with software NOVA 2.0.

Thin films were prepared by spin coating on quartz substrate from dichloromethane solutions. Devices were made by physical vapor deposition (PVD) with a Kenosistec KE500KF vacuum chamber connecting to a glovebox. The base chamber pressure is $5 \times 10^{-7}$ mbar. The chamber includes four effusive (Knudsen) sources and two thermal (Joule) sources. Two microbalances equipped with gold-coated crystal sensor Inficon quartz monitor allow to monitor the thickness of deposited material by the Inficon SQS-242 software. The crystal sensor was calibrated for each material by measuring the thickness of at least three evaporated films by means of a Bruker DektakXT contact profilometer. All the materials were deposited at pressures $\leq 2 \times 10^{-6}$ mbar and deposition rates lower than 1 Å/s. Device were made on ITO-coated glass substrates purchased from KINTEC Company (width of patterned stripe 2.5 mm; thickness, 150 ± 10 nm; sheet resistance 15-20 Ω cm$^{-2}$). ITO substrates were cleaned with detergent and water, sonicated in deionized water, acetone, isopropanol for 15 min each and exposed to an ozone atmosphere for 15 min before loading into the high-vacuum chamber. After organic depositions, LiF and Al layers were deposited by using a mask with 2.5 mm stripe width (placed onto substrates under N$_2$). Each device consists of 4 independent pixels with area of 6.25 mm$^2$. Non-contact AFM images were obtained with a 5500 Keysight system

## ASSOCIATED CONTENT

**Supporting Information**. Synthesis of complex **1** and its characterization, photophysical properties of FIrpic and complex **1,** ESI-MS spectra,, DSCs thermal studies and corresponding HPLC-MS analyses, optical spectrum of FIrpic neat films, device characterization for the unoptimized architecture without Ir(ppz)$_3$ blocking layer, electrical features of the device B$_{80}$, AFM morphological analysis of the FIrpic interface, photoluminescence characterizations of unaged and aged pixels of device A, direct infusion MS and HPLC-MS analysis of unaged and aged devices, calculated absorption spectra of FIrpic and **1.** These materials are supplied as Supporting Information. "This material is available free of charge via the Internet at http://pubs.acs.org."

## AUTHOR INFORMATION

**Corresponding Author**

* Alberto Bossi, email: alberto.bossi@istm.cnr.it, ISTM-CNR via Fantoli 16/15, 20138 Milan, tel: +39 0250995627.
Davide Ceresoli, email: davide.ceresoli@istm.cnr.it, ISTM-CNR via Golgi 19, 20133 Milan, tel: +39 0250314276.
Fausto Cargnoni, email: fausto.cargnoni@istm.cnr.it, ISTM-CNR via Golgi 19, 20133 Milan, tel: +39 0250314272.

**Present Addresses**

†Currently at Department of Chemistry, University of Milan, via Golgi 19, 20133 Milano.

**Author Contributions**

The manuscript was written through contributions of all authors. / All authors have given approval to the final version of the manuscript.

**Funding Sources**

The project has been founded by the Samsung GRO 2014 program "Exciton and polaron induced OLED degradation by combined ab-initio molecular dinamics and experiments" and the progetto Integrato Regione Lombardia and Fondazione CARIPLO (grant numbers 12689/13, 7959/13; Azione 1 e 2, "SmartMatLab centre" and Cariplo Foundation grant 2013-1766).

## ACKNOWLEDGMENT

AB and MP thanks Dr. Alessio Orbelli Biroli and Dr. Daniele Marinotto of ISTM-CNR for the measurements of the film thickness of the organic materials used in OLEDs and Dr. Ivan Andreosso for some preliminary calibrations. Authors greatfully thanks Prof. Gianlorenzo Bussetti, Department of Physics of the Politecnico di Milano, for the AFM investigation and useful discussion. AFM were realized at the Solid-Liquid Interface and Nanomicroscopy (SoLINano) laboratory which is an inter-Departmental lab of the Politecnico di Milano

## REFERENCES

1. Muccini, M.; Toffanin, S. *Organic Light-Emitting Transistors: Towards the Next Generation Display Technology*; Wiley-Science Wise Co-Publication: Hoboken, NJ, 2016.
2. Yersin, H. Highly Efficient OLEDs with Phosphorescent Materials; Wiley-VCH: Weinheim, 2007.
3. Thompson, M. E.; Djurovich, P. E.; Barlow, S.; Marder, S. Organometallic Complexes for Optoelectronic Applications, *Comprehensive Organometallic Chemistry*, ed. D. O'Hare, Elsevier: Oxford, 2007; 12, 101-194.
4. Huckaba, A. J.; Nazeeruddin, M. K. Strategies for Tuning Emission Energy in Phosphorescent Ir(III) Complexes. *Comments Inorg. Chem.* **2016**, *37*, 117-145.
5. Scholz, S.; Kondakov, D.; Lussem, B.; Leo, K. Degradation Mechanisms and Reactions in Organic Light-Emitting Devices. *Chem. Rev.* **2015**, *115*, 8449-8503.
6. Schmidbauer, S.; Hohenleutner, A.; Konig, B. Chemical Degradation in Organic Light-Emitting Devices: Mechanisms and Implications for the Design of New Materials. *Adv. Mater.* **2013**, *25*, 2114-2129.
7. Yamamoto, H.; Weaver, M. S.; Murata, H.; Adachi, C.; Brown, J. J. Understanding extrinsic degradation in phosphorescent OLEDs. *SID Int. Symp. Dig. Tec.* **2014**, *45*, 758–761.
8. Forrest, S. R. Ultrathin Organic Films Grown by Organic Molecular Beam Deposition and Related Techniques. *Chem. Rev.* **1997**, *97*, 1793-1896.
9. Spindler, P.; Hamer, J. W.; Kondakova, M. E. OLED Manufacturing Equipment and Methods, *Handbook of Advanced Lighting Technology*, Springer: Cham, 2017, 417-441.
10. Zhang, Y.; Aziz, H. Degradation Mechanisms in Blue Phosphorescent Organic Light- Emitting Devices by Exciton−Polaron Interactions: Loss in Quantum Yield versus Loss in Charge Balance. *ACS Appl. Mater. Inter.* **2017**, *9*, 636-643.
11. Kim, S.; Bae, H. J.; Park, S.; Kim, W.; Kim, J.; Kim, J. S.; Jung, Y.; Sul, S.; Ihn, S.-G.; Noh, C.; Kim, S.; You Y. Degradation of blue-phosphorescent organic light-emitting devices involves exciton-induced generation of polaron pair within emitting layers, *Nat. Comm.* **2018**, *9*, 1211.
12. Kondakov, D. Y.; Lenhart, W. C.; Nichols, W. F. Operational degradation of organic light-emitting diodes: Mechanism and identification of chemical products. *J. Appl. Phys.* **2007**, *101*, 024512.
13. Sivasubramaniam, V.; Brodkorb, F.; Hanning, S.; Loebl, H. P.; van Elsbergen, V.; Boerner, H.; Scherf, U.; Kreyenschmidt, M. Investigation of FIrpic in PhOLEDs via LC/MS technique. *Cent. Eur. J. Chem.* **2009**, *7*, 836-845.
14. Scholz, S.; Meerheim, R.; Lüssem, B.; Leo, K. Laser desorption/ionization time-of-flight mass spectrometry: A predictive tool for the lifetime of organic light emitting devices. *Appl. Phys. Lett.* **2009**, *94*, 043314.
15. Scholz, S.; Walzer, K.; Leo, K. Analysis of Complete Organic Semiconductor Devices by Laser Desorption/Ionization Time-of-Flight Mass Spectrometry. *Adv. Funct. Mater.* **2008**, *18*, 2541-2547.
16. Jeong, C; Coburn, C.; Idris, M.; Li, Y.; Djurovich, P. I.; Thompson, M. E.; Forrest, S. R. Understanding molecular fragmentation in blue phosphorescent organic light-emitting devices. *Org. Electron.* **2019**, *64*,15–21.
17. Meerheim, R.; Scholz, S.; Olthof, S.; Schwartz, G.; Reineke, S.; Walzer, K.; Leo, K. Influence of charge balance and exciton distribution on efficiency and lifetime of phosphorescent organic light-emitting devices. *J. Appl. Phys.* **2008**, *104*, 014510-014518.
18. de Moraes, I. R.; Scholz, S.; Lussem, B.; Leo, K. Role of oxygen-bonds in the degradation process of phosphorescent organic light emitting diodes. *Appl. Phys. Lett.* **2011**, *99*, 053302-053303.
19. de Moraes, I. R.; Scholz, S.; Leo, K. Influence of the applied charge on the electro-chemical degradation in green phosphorescent organic light emitting diodes. *Org. Electron.* **2016**, *38* 164-171.

20. de Moraes, I. R.; Scholz, S.; Lussem, B.; Leo, K. Analysis of chemical degradation mechanism within sky blue phosphorescent organic light emitting diodes by laser-desorption/ionization time-of-flight mass spectrometry. *Org. Electron.* **2011**, *12*, 341-347.

21. Scholz, S.; Meerheim, R.; Walzer, K.; Leo, K. Chemical degradation mechanisms of organic semiconductor devices. *Proc. SPIE* **2008**, *6999*, 69991B.

22. Ragni, R.; Plummer, E. A.; Brunner, K.; Hofstraat, J. W.; Babudri, F.; Farinola, G. M.; Naso, F.; De Cola, L. Blue emitting iridium complexes: synthesis, photophysics and phosphorescent devices. *J. Mater. Chem.* **2006**, *16*, 1161-1170.

23. Jurow, M. J.; Bossi, A.; Djurovich, P. I.; Thompson, M. E. In Situ Observation of Degradation by Ligand Substitution in Small-Molecule Phosphorescent Organic Light-Emitting Diodes. *Chem. Mater.* **2014**, *26*, 6578-6584.

24. Wang, Q.; Aziz, H. Exciton–Polaron-Induced Aggregation of Organic Electroluminescent Materials: A Major Degradation Mechanism in Wide-Bandgap Phosphorescent and Fluorescent Organic Light-Emitting Devices. *Adv. Optical Mater.* **2015**, *3*, 967-975.

25. Baranoff, E.; Curchod, B. F. E. FIrpic: archetypal blue phosphorescent emitter for electroluminescence. *Dalton Trans.* **2015**, *44*, 8318-8329.

26. Sivasubramaniam, V.; Brodkorb, F.; Hanning, S.; Loebl, H. P.; van Elsbergen, V.; Boerner, H.; Scherf, U.; Kreyenschmidt, M. Fluorine cleavage of the light blue heteroleptic triplet emitter FIrpic. *J. Fluorine Chem.* **2009**, *130*, 640-649.

27. Cazzaniga, M.; Cargnoni, F.; Penconi, M.; Bossi, A.; Ceresoli, D. Unraveling the degradation mechanism in FIrpic based blue OLEDs: I. A theoretical investigation. submitted to *Chem. Mat.* this issues.

28. Sunesh, C. D.; Sunseong, O.; Chandran, M.; Moon, D.; Choe, Y. Effect of ionic liquids on the electroluminescence of yellowish-green light emitting electrochemical cells using bis(2-(2,4-difluorophenyl)pyridine)4,7-diphenyl-1,10-phenanthroline-iridium(III) hexafluorophosphate. *Mat. Chem. Phys.* **2012**, *136*, 173-178.

29. Adamovich, V. I.; Cordero, S. R.; Djurovich, P. I.; Tamayo, A.; Thompson, M. E.; D'Andrade, B. W.; Forrest, S. R. *Org. Electron.* **2003**, *4*, 77-87.

30. Holmes, R. J.; Forrest, S. R.; Tung, Y. J.; Kwong, R. C.; Brown, J. J.; Garon, S.; Thompson, M. E. Blue organic electrophosphorescence using exothermic host–guest energy transfer. *Appl. Phys. Lett.* **2003**, *82*, 2422-2424.

31. Aziz, H.; Popovic, Z.; Tripp, C. P.; Hu, N.-X.; Hor, A.-M.; Xu, G. Degradation processes at the cathode/organic interface in organic light emitting devices with Mg:Ag cathodes. *Appl. Phys. Lett.* **1998**, *72*, 2642-2644.

32. de Moraes, I. R.; Scholz, S.; Lussem, B.; Leo, K. Chemical degradation processes of highly stable red phosphorescent organic light emitting diodes. *Org. Electron.* **2012**, *13*, 1900-1907.

33. In an oversimplified scheme, approximately the amount of **1** in device B, based on the volume of the pixel emitting layer and considering a reasonable estimate of the lower limit for FIrpic degradation of about 5% of molecules, yields less than $5 \times 10^{-8}$ M (0.03 µg/ml) of complex **1**.

34. Note: in ACN-$H_2$O reverse phase HPLC, *Fac-Mer* mixture of tris-cyclometalated Ir complexes (homo or heteroleptic), the *Mer* isomer is eluted a shorted time than the corresponding *Fac* one.

SYNOPSIS TOC Exciton induced degradation of FIrpic emitter in Blue-OLEDs. Paramount importance of addressing the chemical stability of the molecules both *ex-situ* and *in-situ*.

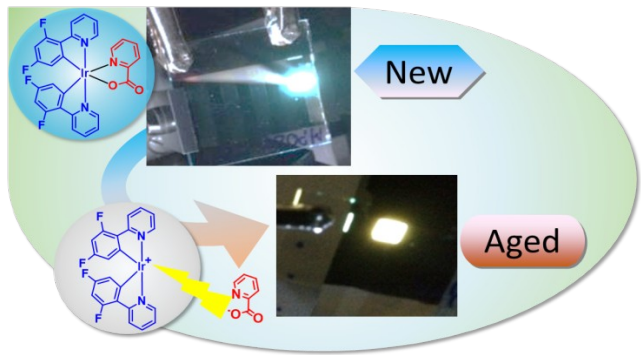